\documentclass{article}

\usepackage{arxiv}

\usepackage[utf8]{inputenc} % allow utf-8 input
\usepackage[T1]{fontenc}    % use 8-bit T1 fonts
\usepackage{hyperref}       % hyperlinks
\usepackage{url}            % simple URL typesetting
\usepackage{mathrsfs}
\usepackage{amsmath}
\usepackage{graphicx}
\usepackage{cite}
\usepackage{caption}
\usepackage{amssymb}
\usepackage{algorithm}
\usepackage[noend]{algpseudocode}
\usepackage[short]{optidef}
\usepackage{subcaption}
\usepackage{svg}

\title{Autoregressive models for biomedical signal processing}

\author{Jonas F.\ Haderlein\thanks{This research was conducted in the Australian Research Council Training Centre in Cognitive Computing for Medical Technologies (project number ICI70200030) and funded by the Australian Government.},~Andre D.\ H.\ Peterson%
\thanks{A.D.H.\ Peterson and D.B.\ Grayden are with the Department of Medicine, St Vincent's Hospital, The University of Melbourne.},
\AND Anthony N.\ Burkitt,~Iven M.\ Y.\ Mareels\thanks{I.\ Mareels is with the Institute for Innovation, Science and Sustainability, Federation University Australia, Mt Helen, Vic 3350},~David B.\ Grayden% <-this % stops a space
\thanks{J.\ Haderlein, A.D.H.\ Peterson, A.N.\ Burkitt and D.B.\ Grayden are with the Department of Biomedical Engineering, University of Melbourne VIC 3010, Australia.}% <-this % stops a sp
\thanks{D.B.\ Grayden is with the Graeme Clark Institute, University of Melbourne VIC 3010, Australia.}% <-this % stops a sp
}

\begin{document}

\maketitle
%\thispagestyle{empty}
%\pagestyle{empty}

%%%%%%%%%%%%%%%%%%%%%%%%%%%%%%%%%%%%%%%%%%%%%%%%%%%%%%%%%%%%%%%%%%%%%%%%%%%%%%%%
\begin{abstract}

Autoregressive models are ubiquitous tools for the analysis of time series in many domains such as computational neuroscience and biomedical engineering. 
In these domains, data is, for example, collected from measurements of brain activity.
Crucially, this data is subject to measurement errors as well as uncertainties in the underlying system model.
As a result, standard signal processing using autoregressive model estimators may be biased. 
We present a framework for autoregressive modelling that incorporates these uncertainties explicitly via an overparameterised loss function. 
To optimise this loss, we derive an algorithm that alternates between state and parameter estimation.
Our work shows that the procedure is able to successfully denoise time series and successfully reconstruct system parameters.
This new paradigm can be used in a multitude of applications in neuroscience such as brain-computer interface data analysis and better understanding of brain dynamics in diseases such as epilepsy.
\end{abstract}

%%%%%%%%%%%%%%%%%%%%%%%%%%%%%%%%%%%%%%%%%%%%%%%%%%%%%%%%%%%%%%%%%%%%%%%%%%%%%%%%
\section{Introduction}
Biomedical signal processing necessitates a principled treatment of uncertainties in the data generating process and measurement acquisition.
Recent applications in this field are the analyses of electroencephalogram (EEG) or other brain-computer interface (BCI) measurements. 
Careful data analysis is crucial for a study of brain function and dysfunction, brain signal decoding, and BCI-based control of external devices in fields like bionic prostheses and robotics.
However, inherent sources of uncertainty are, inevitably, measurement noise as well as the underlying system model.
The latter source is a key challenge in the biomedical domain, since system models (usually in the forms of dynamical systems) for the brain or the observed neuronal populations are unknown, and the system state is inherently immeasurable with current technology.

% PAragraph on the use of AR models
Instead of postulating data-generating models for the brain, many approaches focus on black-box analysis with autoregressive (AR) models.
Due to the time-varying and non-autonomous nature of the true dynamics, AR models are commonly fitted over short finite horizons of measurements that are oscillatory in nature. 
In the neuroscience domain, AR models are used to infer neuronal network structures \cite{Greenblatt2012, Shah2019, Li2021b} and for intervention or control mechanisms \cite{Brogin2023}. 
In the BCI domain, AR models are often used to estimate the frequency spectrum of brain activity \cite{Wolpaw2012, Meng2016, Padfield2019}. 
We observe that it is the reconstruction of parameters and denoised states that are often of interest here, rather than time series prediction.
As we show in the following, classical estimators for AR models can be biased under noise, and may perform sub-optimally in such tasks.

% Paragraph on AR sysid and porblems
AR modelling is an area within system identification, a well-studied and broad field \cite{Ljung1987, Billings2013a, PintelonSchoukens}.
The general problem is to model the dynamics of a partially observed system from measurements only.
Denote the measurements $y_t \in R^p$ with finite time index  $t=1,\cdots,N$.
In AR modelling, a function $g$ is proposed that models the evolution of $y_t$ as an approximation of the true dynamics $f$ with an unknown state dimension. 
The existence of such models $g$ follows from the Takens-Aeyels-Sauer theorem \cite{Takens1981} (having observed noiseless measurements of sufficient length $N$) with a delay embedding of the form
\begin{equation}
\label{embedding}
\begin{array}{lcll}
y_{t+1} &=& g_{\Theta}(y_t, y_{t-1},\cdots,y_{t-r+1})^T, ~~0 < r < N.
\end{array}
\end{equation}
Here, $\Theta$ is a parameter vector specifying the model and $r$ reflects our prior assumption of the true system state dimension. 

Linear or nonlinear autoregressive (NAR) models $g_{\Theta}$ can be constructed with basis functions of various form, such as linear or polynomial functions or via (recurrent) neural networks \cite{Billings2013a}. 
Modelling, i.e., system identification, is then a search for optimal model parameters $\Theta$.

In the following, we assume $p = 1$ for simplicity, but the analysis can easily be generalised for $p > 1$, as shown in Example \ref{example3}.

The following sections provide an overview of the two main modelling frameworks \cite{Schoukens2019}.

\subsection{Prediction error framework}
\label{PredError}

The prediction error framework assumes a model
\begin{equation}
\begin{array}{lcll}
y_{t+1} &=& g_{\Theta}(y_t, y_{t-1},\cdots,y_{t-r+1})+\nu_t,
\end{array}
\end{equation}
where $\nu_t$ is a transition noise. 

One optimises the one-step-ahead prediction loss $\ell_P(r, \Theta, Y)$ over parameters $\{r, \Theta \}$:
\begin{mini}
  {r, \Theta}{\sum_{t=r}^{N-1} \left(y_{t+1} - g_{\Theta}({x}_t) \right)^2}{}{},
 \label{predictionloss} 
 \end{mini}
with delay embedding $x_{t} = (y_t, y_{t-1},\cdots,y_{t-r+1})^T \in R^r$ and measurements $Y = (y_1, \cdots, y_N)^T
$. 
%As such, this framework is best suited for short-term predictions of the system.

By directly using $Y$ as input and target variables, this approach assigns a high degree of trust in the measurements and, as a result, reconstructs biased $\Theta$ in the presence of measurement noise \cite{Piroddi2003, Aguirre2010, Schoukens2019}.

The linear AR case can be solved directly with least squares techniques, which can be readily extended to any linear-in-the-parameters model class \cite{Homer2004a, Tran2017a, Brunton2016b}. We note that Koopman theory also provides a way to model nonlinear behaviour in terms of measurements only \cite{Iacob2021, Brunton2021, Sznaier2021}. 
%In the presence of noise (sometimes called `errors-in-variables'), \textit{orthogonal least squares} may be applied, which neglects information about the distribution of $y_t$ (see e.g. \cite{Voss2004} and references therein).
Alternative methods that reconstruct linear state-space models from measurements are known as sub-space estimation \cite{VanOverschee1993, Ljung1987}.

Examples of nonlinear models based on the program in (\ref{predictionloss}) have been presented in \cite{Lu2019, Ljung2020, Ljung2020a}.

\subsection{Simulation error framework}
\label{SimError}

The simulation error framework asks for a model
\begin{equation}
\begin{array}{lcll}
\hat{y}_{t+1} &=& g_{\Theta}(\hat{y}_t, \hat{y}_{t-1},\cdots,\hat{y}_{t-r+1}),\\
y_t &=& \hat{y}_t+\mu_t,
\end{array}
\end{equation}
where $\mu_t$ is measurement noise. 
We optimise the simulation loss $\ell_S(r, \Theta, \hat{x}_{r}, Y)$ over parameters $\{r, \Theta, \hat{x}_r \}$, including the initial condition $\hat{x}_r = (\hat{y}_{r}, \cdots, \hat{y}_1)^T$:
\begin{mini}
  {r, \Theta, \hat{x}_{r}}{\sum_{t=r+1}^{N}(y_t - \hat{y}_t)^2}{}{}
  \addConstraint{\hat{y}_{t+1}}{= g_{\Theta}(\hat{x}_t),\quad}{t=r, \cdots, N-1},
 \label{simulationerror} \end{mini}
where $\hat{y}_t$ is the simulation estimate of the model at time $t$ and $\hat{x}_t = (\hat{y}_{t}, \cdots, \hat{y}_{t-r+1} )^T$ is the delay embedding.

This optimisation provides a denoised state estimate $\hat{y}_t ~\forall t$, but, at the same time, assumes no uncertainty in the model. While in principle appealing, it is often complex to optimise the non-convex $\ell_S$ in practice, even for very simple $g_{\Theta}$  \cite{Voss2004}.

The complexity of the loss in (\ref{simulationerror}) motivates the use of advanced optimisation programs such as multiple shooting or evolutionary algorithms \cite{Zhang2004a,Decuyper2020b,  Aguirre2010}.
Alternatively, prediction error-based models can be further pruned by minimising a simulation error loss \cite{Piroddi2003}.
Decomposing the trajectory $Y$ and optimising only over multi-step predictions is also employed in order to relax the problem \cite{farina2011, Forgione2020}.

Recent work includes the use of neural delay differential equations, a class of neural network basis functions that can be tuned via simulation error optimisation \cite{Schlaginhaufen2021}.

\subsection{Outline}
Neither prediction error nor simulation error frameworks incorporate both measurement and model errors separately. 
Alternative approaches that do so have been proposed for known models \cite{Mareels2002, Haderlein1} and black-box input-output models \cite{VanGorp2000a}.
Similarly, model estimation in Bayesian frameworks deals with both kinds of uncertainties and is widely adopted \cite{West1996, Pole2018, Gelman2020}.
One approach for these methods is expectation-maximisation, which requires iterative state and parameter estimation, e.g., via Kalman filtering  \cite{Shumway1982, Ghahramani1999, Roweis1999}.

Our main contribution in this study is to extend the classical frameworks by incorporating these ideas into AR modelling.
%We present an estimation technique for AR modelling in which both state and measurement noise are accounted for. 
We first derive an overparameterised loss function so that we neither only search for a denoised state with total certainty in the model (simulation error framework) nor totally rely on the measurements (prediction error framework).
We then propose a novel optimisation algorithm akin to expectation-maximisation.
This recipe solves both parameter and state estimation in batch mode, as opposed to recursive techniques such as Kalman filtering and smoothing \cite{Ljung1987}.

\section{AR models for state and measurement errors}
\label{method}

%The above approaches are particular instances of finding minima of a loss function that directly follows from the log-likelihood over states and parameters. 
With an AR model in state-space form,
\begin{equation}
\begin{array}{lcll}
\hat{x}_{t+1} &=& g_{\Theta}(\hat{x}_t) + \nu_t,\\
y_t &=& C\hat{x}_t+\mu_t,
\label{NARstatespace}
\end{array}
\end{equation}
with row vector $C \in R^{1 \times r}$, we minimise a loss, over the set $\{r, \hat{\Theta}, \hat{X}\}$ derived from the log-likelihood of state transitions \cite{Roweis1999},
\begin{equation}
\begin{split}
\label{loss}
\ell(r, \Theta, \hat{X}, Y) = &\sum_{t=r}^{N-1} \left(\hat{x}_{t+1} - g_{\Theta}(\hat{x}_t) \right)^2 + \\
&\rho \sum_{t=r}^{N}(y_t - C\hat{x}_t)^2,
\end{split}
\end{equation}
with $\rho > 0$ as a variable expressing our belief in the measurements vs. the model and $\hat{x}_t$ as a state estimate of respective delays. 

The loss in (\ref{loss}) incorporates both denoised states $\hat{X} = (\hat{x}_r, \cdots \hat{x}_N)^T$ and parameters $\Theta$ that need to be optimised. It is thus, in some sense, a relaxation of problem (\ref{simulationerror}) via the auxiliary state variables. 

We discuss the linear and nonlinear case in the following.

\subsection{The linear finite horizon case}

We consider the linear map
\begin{equation}
\label{takensembedding}
\begin{array}{lcll}
   \hat{x}_{t+1} &=& A\hat{x}_{t}+\nu_t,\\
   y_t &=& C\hat{x}_{t}+\mu_t,
   \end{array}
\end{equation}
with a priori unknown $A$ in companion matrix form with the autoregressive coefficients $\Theta \in R^r$ in the first column and trivial $C = (1, 0, \cdots, 0)$. 

In the prediction error framework, $\Theta$ is approximated by a (regularised) least squares estimate,
\begin{equation}
\label{linearparameter}
\hat{\Theta} = (\Gamma^T\Gamma + \lambda I)^{-1} \Gamma^T {Y}_{+},
\end{equation}
where $\Gamma = (\hat{x}_{r}, \hat{x}_{r+1}, \cdots, \hat{x}_{N-1})^T$ is a matrix of the respective delay embeddings, ${Y}_{+} = (\hat{y}_{r+1}, \hat{y}_{r+2}, \cdots, \hat{y}_N)^T$, and $\hat{Y} = Y$, $\hat{X} = X$. $\lambda \geq 0$ is the regularisation strength and $I$ is an identity matrix of appropriate size. 

We assume throughout that the above inverse exists, for which we require sufficiently large $N$.

Given $C$ as well as estimates for $\hat{A}, \hat{Y}$, we can then map all available information onto an updated state estimate, 
\begin{equation}
\label{linearestimator}
    \hat{X} =  \mathcal{O}^{-1} \rho \mathcal{C}^T\hat{Y},
\end{equation}
where $\hat{X}=(\hat{x}_1^T \cdots \hat{x}_N^T)^T$ and matrices $\mathcal{O}, \mathcal{C}$ for state reconstruction as defined in \cite{Haderlein1}. 

Thus, we propose the alternating minimisation of the loss in (\ref{loss}) outlined in Algorithm~\ref{alg1}, which does not require the use of first-order or second-order optimisation methods.

\def\NoNumber#1{{\def\alglinenumber##1{}\State #1}\addtocounter{ALG@line}{-1}}
\begin{algorithm}[H]
	\caption{AR model identification (given $r, Y, C, \lambda, \rho$)}
	\label{alg1}
	\begin{algorithmic}[1]
	\State initialise $\hat{Y} = Y$ and $\hat{Y}_+ = Y_+$
	\While {$\hat{A}, \hat{Y}$ not converged}
		\State Based on the current estimate $\hat{Y}$, find a new $\hat{A}$ by: $$\hat{\Theta} = (\Gamma^T\Gamma + \lambda I)^{-1} \Gamma^T \hat{Y}_{+}$$
		\State Based on $\hat{A}, C$ calculate $\mathcal{O}$
		\State Retrieve a new state estimate based on $Y$: $$\hat{X} =  \mathcal{O}^{-1} \rho \mathcal{C}^TY$$
		\State Retrieve updated $\hat{Y}, \hat{Y}_+$ from $\hat{X}$:
		$$\hat{Y} = \mathcal{C} \hat{X}$$
		\EndWhile
		
	\end{algorithmic} 
\end{algorithm}

The convergence of Algorithm~1 (here for $\lambda = 0$) can be argued as follows. We can associate a loss with Algorithm~1, 
\begin{equation}
\label{linloss}
\ell(\hat{A}, \hat{X}, Y) =  \sum_{t=r}^{N-1} \left(\hat{x}_{t+1} - \hat{A} \hat{x}_t \right)^2 + \rho \sum_{t=r}^{N}(y_t - C\hat{x}_t)^2,
\end{equation}
which is bounded from below by zero. 
The parameter estimation step results in a new estimate $\hat{\Theta}$
at fixed $\hat{Y},\hat{X}$. Being the least-squares estimate, it hence minimises the contribution of the first term to the loss, while keeping the second term unchanged. 
The state estimation step, resulting in a new $\hat{X}$, finds exactly the $\hat{X}$ that minimises $\ell$ for a given $\hat{A}$. Loss $\ell$ in (\ref{linloss}) is thus a Lyapunov function for the algorithm, which has to converge as a result.

\subsection{The nonlinear finite horizon case}

We consider further the case of nonlinear AR models that are nonetheless linear-in-the-parameters. 
These are models  with basis functions such as polynomials \cite{Decuyper2020, Brunton2016b} and path signatures from rough path theory that have been used in time series applications \cite{Lyons1998, Levin2013, Lyons2014, Chevyrev2021}. 
With a vector of basis functions $s_t$ of the embedding $x_t$, we postulate the state-space map
\begin{equation}
\label{signaturemodel}
\begin{array}{lcll}
   s_{t+1} &=& A s_t+\nu_t,\\
   y_t &=& Cs_{t}+\mu_t.
   \end{array}
\end{equation}
$s_t$ could, for example, be the vector of all polynomials up to a certain order (of the delays $y_r, \cdots, y_{t-r+1}$) that evolves under the dynamics $A$.

Here, much like in the linear dynamics case, we can find estimates $\hat{A}$ and $\hat{C}$ by least squares for any $\hat{Y}$ and auxiliary matrices $\Gamma_- = (s_{r}, s_{r+1}, \cdots, s_{N-1})^T$ and $\Gamma_+ = (s_{r+1}, s_{r+2}, \cdots, s_{N})^T$ of stacked basis functions. 
Note that, here, C is not necessarily trivial but defined by a vector that maps the state $s_{t}$ back to the measurement $y_t$.

We present a NAR modelling approach with the signature up to a certain depth defining a state-space embedding. 
A vector of signatures $s_t = s_t^d$ of depth $d$ is extracted from a geometric path of $\hat{x}_t$ in $R^{r \times k}, k>1$, i.e.,
\begin{equation}
\begin{array}{lcll}
   s^d_{t} &=& \text{Sig}^d(\hat{x}_t) \in R^{(k^{d+1}-1)/(k-1)}.
   \end{array}
\end{equation}
We can approximate a nonlinear $g$ with a linear function in signature space, provided large enough $d$, so that
\begin{equation}
\begin{array}{lcll}
    \| g_{\Theta}(\hat{x}_t) - A(s^d_{t}) \| \leq \epsilon,
   \end{array}
\end{equation}
for any $\epsilon > 0$ \cite{Levin2013}. Thus, the linear-in-the-parameters signature model in (\ref{signaturemodel}) can represent the nonlinear case (\ref{NARstatespace}).

The NAR procedure is outlined in Algorithm~\ref{alg2}. 

\def\NoNumber#1{{\def\alglinenumber##1{}\State #1}\addtocounter{ALG@line}{-1}}
\begin{algorithm}[H]
	\caption{NAR model identification (given $r, d, Y, \lambda, \rho$)}
	\label{alg2}
	\begin{algorithmic}[1]
	\State initialise $\hat{Y} = Y$ and $\hat{Y}_+ = Y_+$
	\While {$\hat{A}, \hat{C}, \hat{Y}$ not converged}
	    \State Compute all $s_t^d$ from $\hat{Y}$ and form $\Gamma_-, \Gamma_+$
		\State Based on the current estimate $\hat{Y}$, find a new $\hat{A}$: $$\hat{A} = (\Gamma_-^T\Gamma_- + \lambda I)^{-1} \Gamma_-^T \Gamma_+$$
		\State Based on the current estimate $\hat{Y}$, calculate: $$\hat{C} = (\Gamma_+^T\Gamma_+ + \lambda I)^{-1} \Gamma_+^T \hat{Y}_{+}$$
		\State Based on updated $\hat{A}, \hat{C}$ calculate $\mathcal{O}$ and $\mathcal{C}$
		\State Retrieve a new state estimate based on $Y$: $$\hat{X} =  (\mathcal{O}+ \lambda I)^{-1} \rho \mathcal{C}^TY$$
		\State Retrieve updated $~\hat{Y} = \mathcal{C} \hat{X}$
		\EndWhile
		
	\end{algorithmic} 
\end{algorithm}

\section{Examples}

\subsection{Finding the state dimension $r$}

Many standard techniques exist for the estimation of an approximate state dimension $\hat{r}$. We aim to show that, using our methodology, the following measures provide a good estimate for marginally stable systems like EEG: a) the value of the normalised loss function (\ref{linloss}), i.e., $\frac{1}{N} \ell$, and b) the smallest eigenvalue (in magnitude) of the reconstructed system matrix $\hat{A}$, since, for modelling, all eigenvalues are ideally close to the unit disc.

We simulate a stable AR model $x_{t+1}=A x_t + \nu_t,~ y_t = C x_t + \mu_t$ with $r_{\mathrm{true}}=5$, where $A$ is a companion matrix with five eigenvalues of magnitude 1 (the angles of the eigenvalues left of the imaginary axis are $3\pi/5 + i~\pi/5,~i=0,1,\cdots,4 $), $C=(1,0, \cdots,0)$, $x_1=(1,0,\cdots,0)^T$, $N=200$, and both state and measurement noise i.i.d., $\nu_t \sim N(0, 0.01)$ and $\mu_t \sim N(0, 1)$. 
We simulate 100 trajectories, and evaluate the above criteria for $\hat{r} = 1, \cdots, 10$ using Algorithm~1 for 100 iterations each, with $\rho = 0.1$ and $\lambda = 0$.

\begin{figure}[]
     \centering
     \begin{subfigure}{0.6\columnwidth}
    \includegraphics[width=\columnwidth]{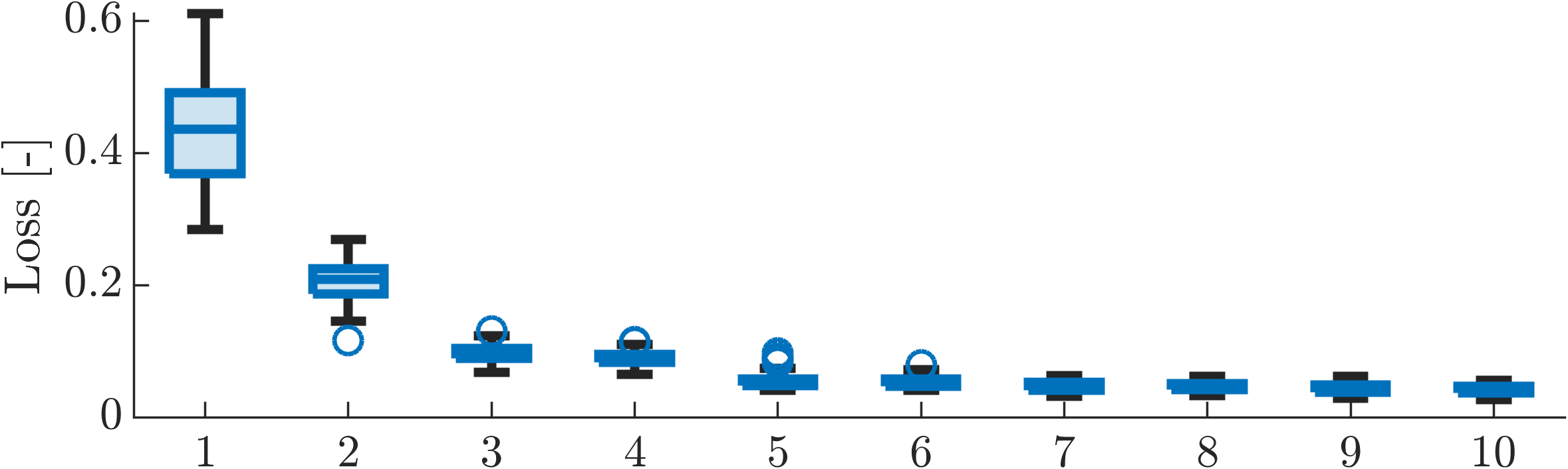}
         \caption{}
     \end{subfigure}
     \begin{subfigure}{0.6\columnwidth}
    \includegraphics[width=\columnwidth]{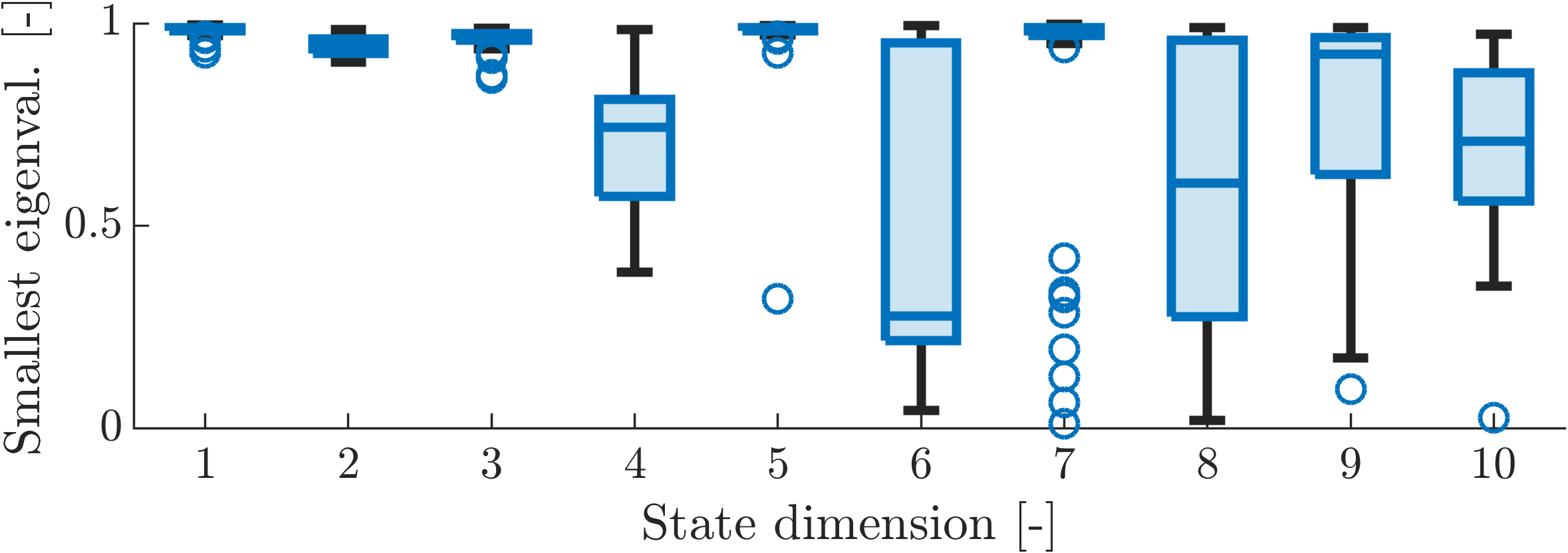}
         \caption{}
     \end{subfigure}
    \caption{Boxplots of (a) normalised loss, and (b) smallest $\hat{A}$ eigenvalue over state dimensions $\hat{r} = 1, \cdots, 10$, after 100 iterations of Algorithm~1, for 100 simulations with $r=5$.}
    \label{figureResDim}
\end{figure}

Fig.~\ref{figureResDim} confirms that the loss $\ell$ decreases with increasing $\hat{r}$, but plateaus for $\hat{r} > r_{\mathrm{true}}$. We observe additionally that the minimum eigenvalue of the resulting estimate $\hat{A}$ has a peak at the correct embedding size $\hat{r} = r_{\mathrm{true}}$. 
The latter property holds only after multiple iterations of Algorithm~1: in this example, the standard least squares estimate (i.e., Algorithm~1 after one iteration) gives, in the median, a smallest $\hat{A}$ eigenvalue of $0.514$, whereas, after 100 iterations, the median of the smallest eigenvalue is $0.987$ (see Fig.~\ref{figureResDim} at $\hat{r}=5$). This shows that the algorithm  converges towards stable dynamics, as in the simulation, with its eigenvalues closer to the unit disc. This may help the estimation of model order $\hat{r}$, a  crucial step in AR modelling.

\subsection{Convergence under noise}

AR modelling has, in general, no unique solution. Algorithm 1 may thus not always converge to a single global optimum, even in controlled cases with simulated data. 

In order to analyse the convergence properties of Algorithm~1, we test the alternating minimisation towards ground truth values from the same model, $x_{t+1}=A x_t + \nu_t,~ y_t = C x_t + \mu_t$ with $r=5$ ($A$, $C$, $x_1$, and $N$ same as above), and same random i.i.d. noise, $\nu_t \sim N(0, 0.01)$ and $\mu_t \sim N(0, 1)$. 
We simulate 100 trajectories, and run Algorithm~1 for 100 iterations with $\rho = 0.1$ and $\lambda = 0$.

\begin{figure}[]
     \centering
     \begin{subfigure}{0.6\columnwidth}
    \includegraphics[width=\columnwidth]{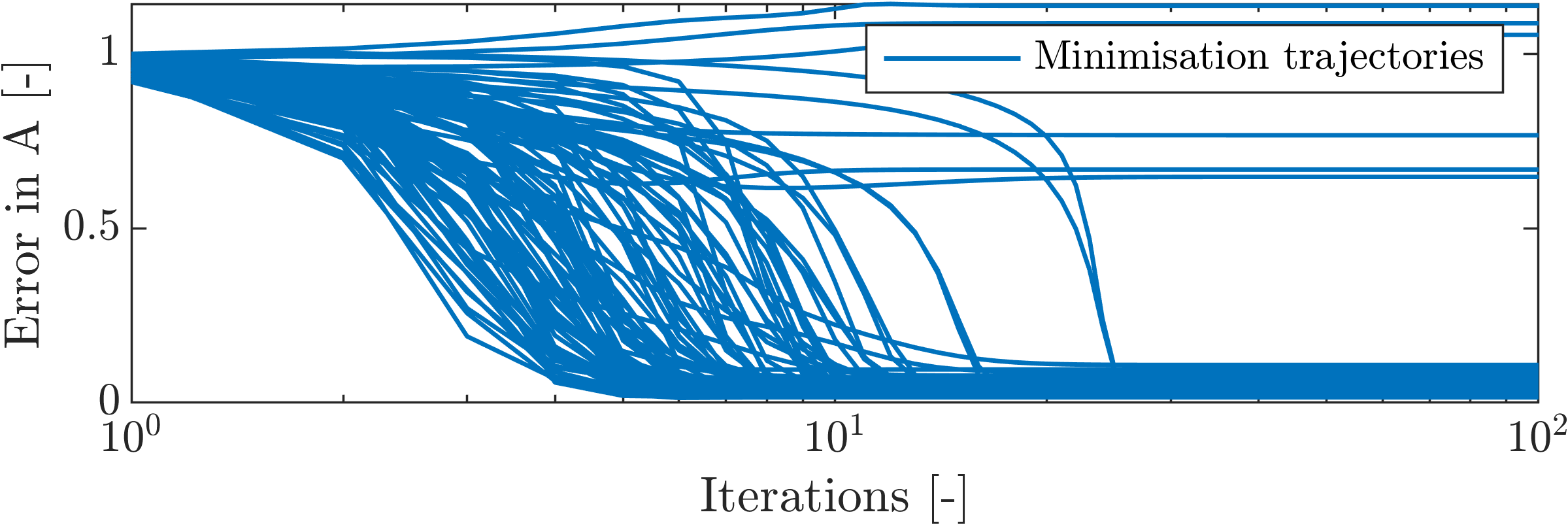}
         \caption{}
     \end{subfigure}
     \begin{subfigure}{0.6\columnwidth}
    \includegraphics[width=\columnwidth]{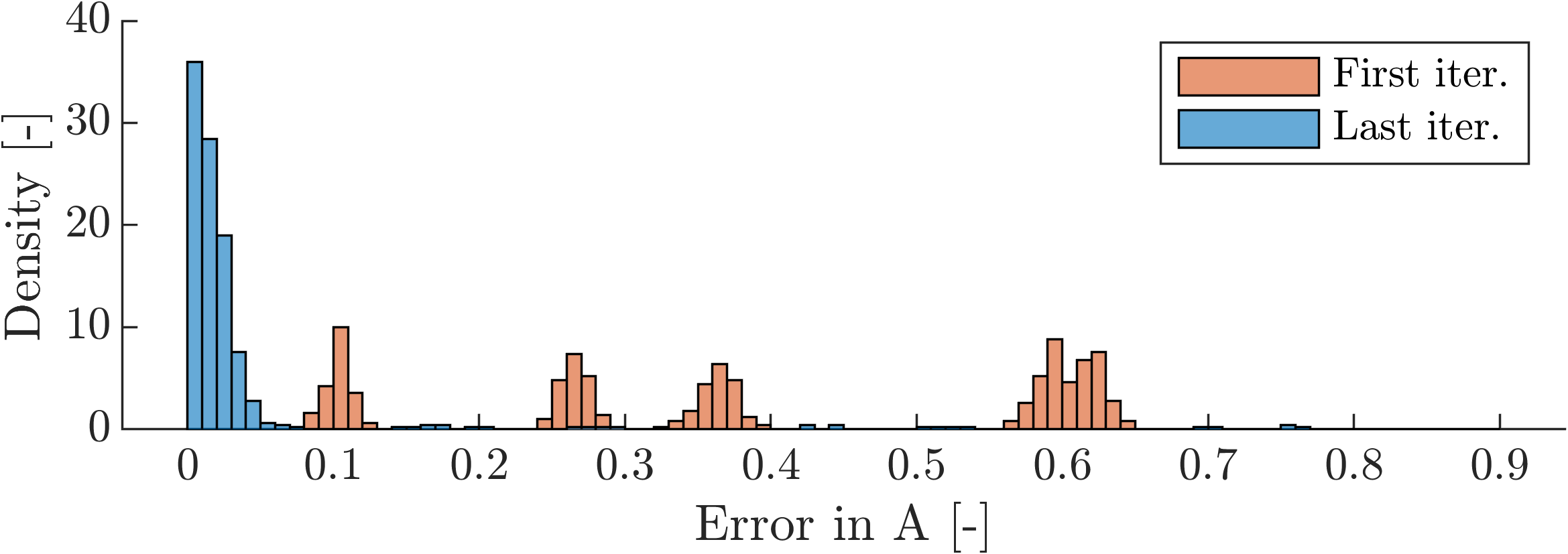}
         \caption{}
     \end{subfigure}
     \begin{subfigure}{0.6\columnwidth}
    \includegraphics[width=\columnwidth]{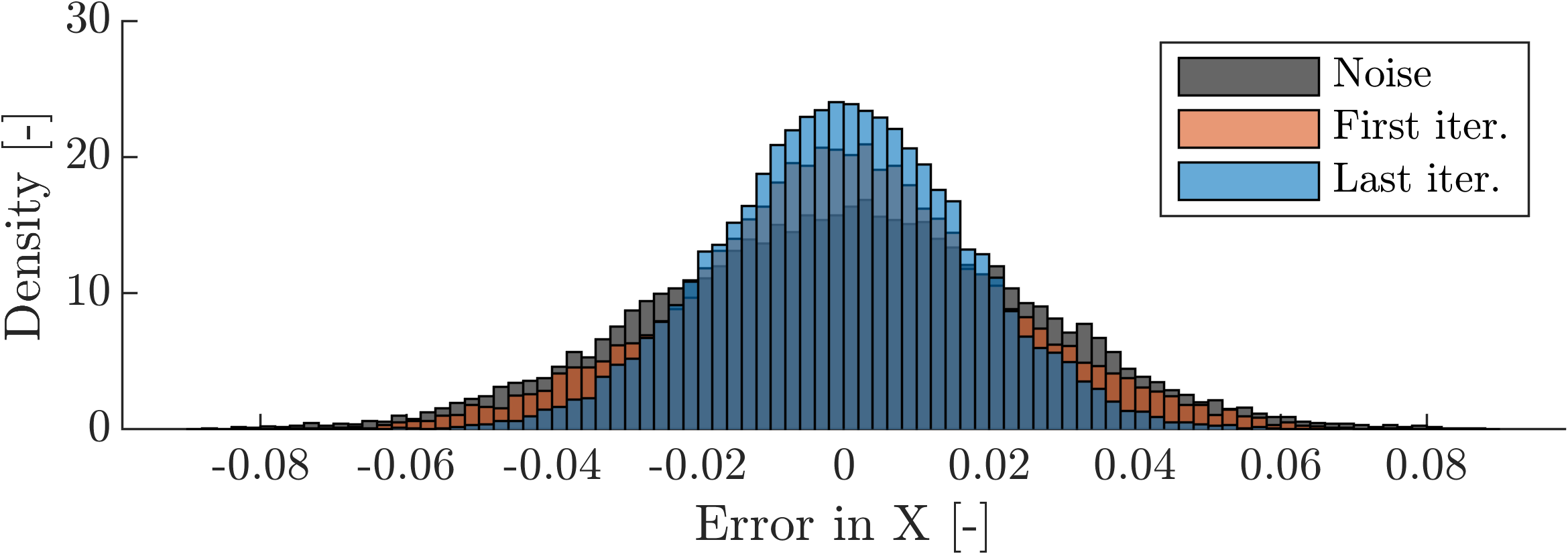}
         \caption{}
     \end{subfigure}
    \caption{Algorithm~1 over 100 iterations: a) Convergence to ground truth $A$ as per error norm $e_{\|\Theta\|} = \|\hat{\Theta} - \Theta\|/ \|\Theta\|$; b) Error distribution in AR coefficients $e_{\Theta} = (\hat{\Theta} - \Theta)/ \|\Theta\|$ for first and last iteration; c) Error distribution in state reconstruction $e_X = (\hat{Y} - x)/ \|x\|$, for measurements $\hat{Y} = Y$ (black), $\hat{Y}$ after 1 iteration (red), and $\hat{Y}$ after 100 iterations (blue). Iteration 1 is the least squares estimate for $A$. }
    \label{figureResConv}
\end{figure}

We analyse the error norm $e_{\|\Theta\|} = \|\hat{\Theta} - \Theta\|/ \|\Theta\|$ and relative deviation $e_{\Theta} = (\hat{\Theta} - \Theta)/ \|\Theta\|$ in reconstructing an estimate $\hat{\Theta}$, as well as the error $e_x = \|\hat{Y} - x\|/ \|x\|$ of the denoised estimate $\hat{Y}$ vs. the true noiseless trajectory $x = (Cx_1, \cdots, Cx_t)^T$.
Results are depicted in Fig.~\ref{figureResConv}, with 6 out of 100 iterations not converging to the ground truth $A$ (despite small loss, compare with Fig. \ref{figureResDim}), but a different local minimum solution.

\subsection{AR EEG connectivity}
\label{example3}

\begin{figure}[]
      \centering
      \includegraphics[width=0.6\columnwidth]{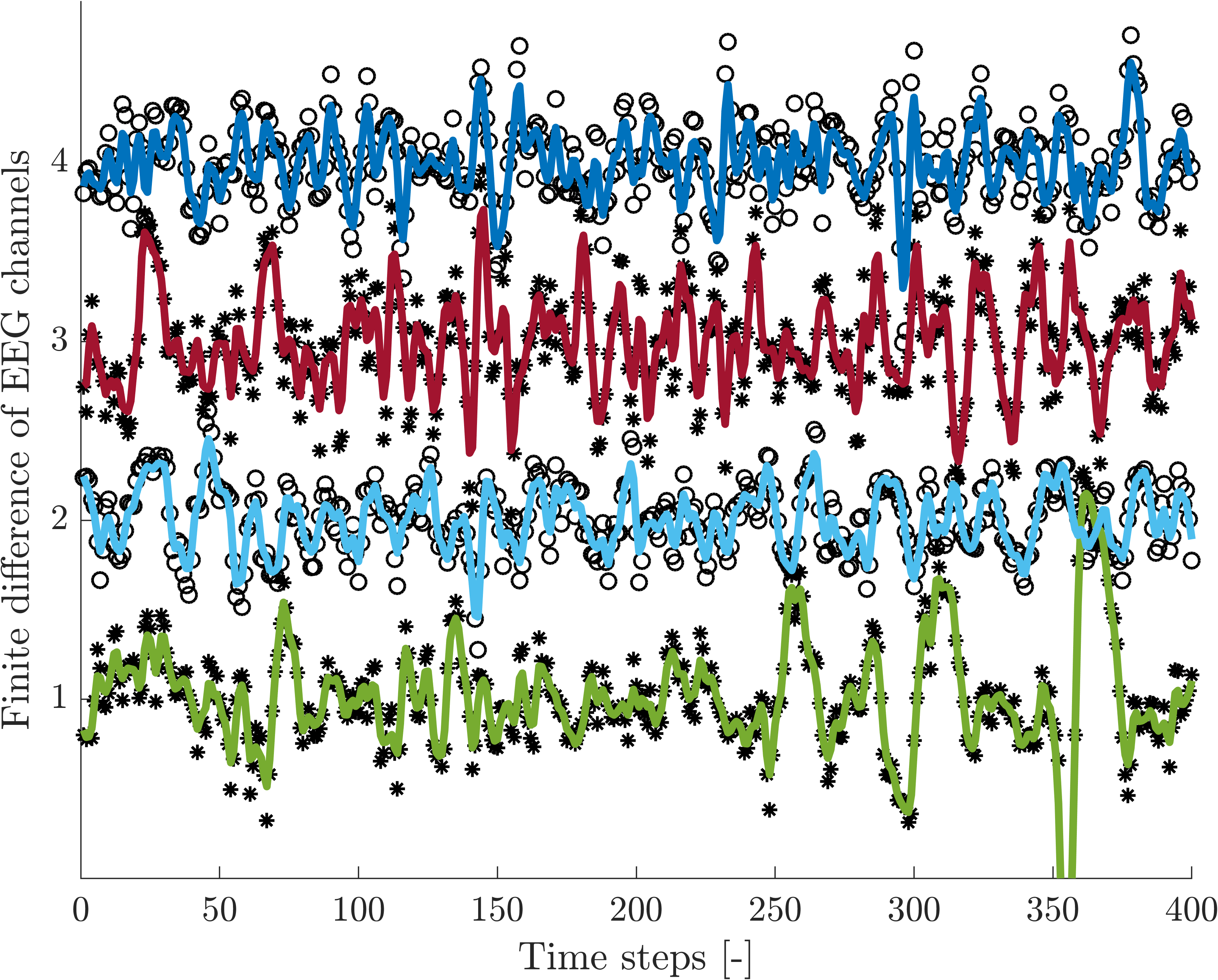}
      \caption{Qualitative plot of the first-order difference in EEG channels, $Y$, (black) and the denoised estimate $\hat{Y}$ for each channel (coloured lines) after 100 iterations of Algorithm~1.}
      \label{figureResSeiz}
\end{figure}

We employ a version of Algorithm~1 for the estimation of linear EEG inter-channel connectivity (like cross-correlation). To this end, we use seizure~2 and (adjacent) channels~$1-4$ from intracranial EEG data available online \cite{Karoly2018}. We take only the first-order finite difference in EEG value of subsequent time steps in order to remove the high autocorrelation, low frequency part of the signal.
Take a time horizon of the first 400 time steps, corresponding to one second of EEG. Denote this vector $Y = (y_1, \cdots, y_{400})^T \in R^{400 \times 4}$. See Figure~\ref{figureResSeiz} for a plot of these measurements.

We fit a four-dimensional AR model of order 1, i.e., $r=1,~p=4$. Thus, our model for this data is $x_{t+1}=A x_t + \nu_t,~ y_t = C x_t + \mu_t, ~ t=1, \cdots, 400,$ where 
$x_t = y_t \in R^4$ are measurements at time step $t$ only.
$\hat{A}$ then contains the $4 \times 4$ connectivity terms and $C \in R^{4 \times 4}$ is an identity matrix. 

\begin{figure}[]
     \centering
     \begin{subfigure}{0.3\columnwidth}
    \includegraphics[width=\columnwidth]{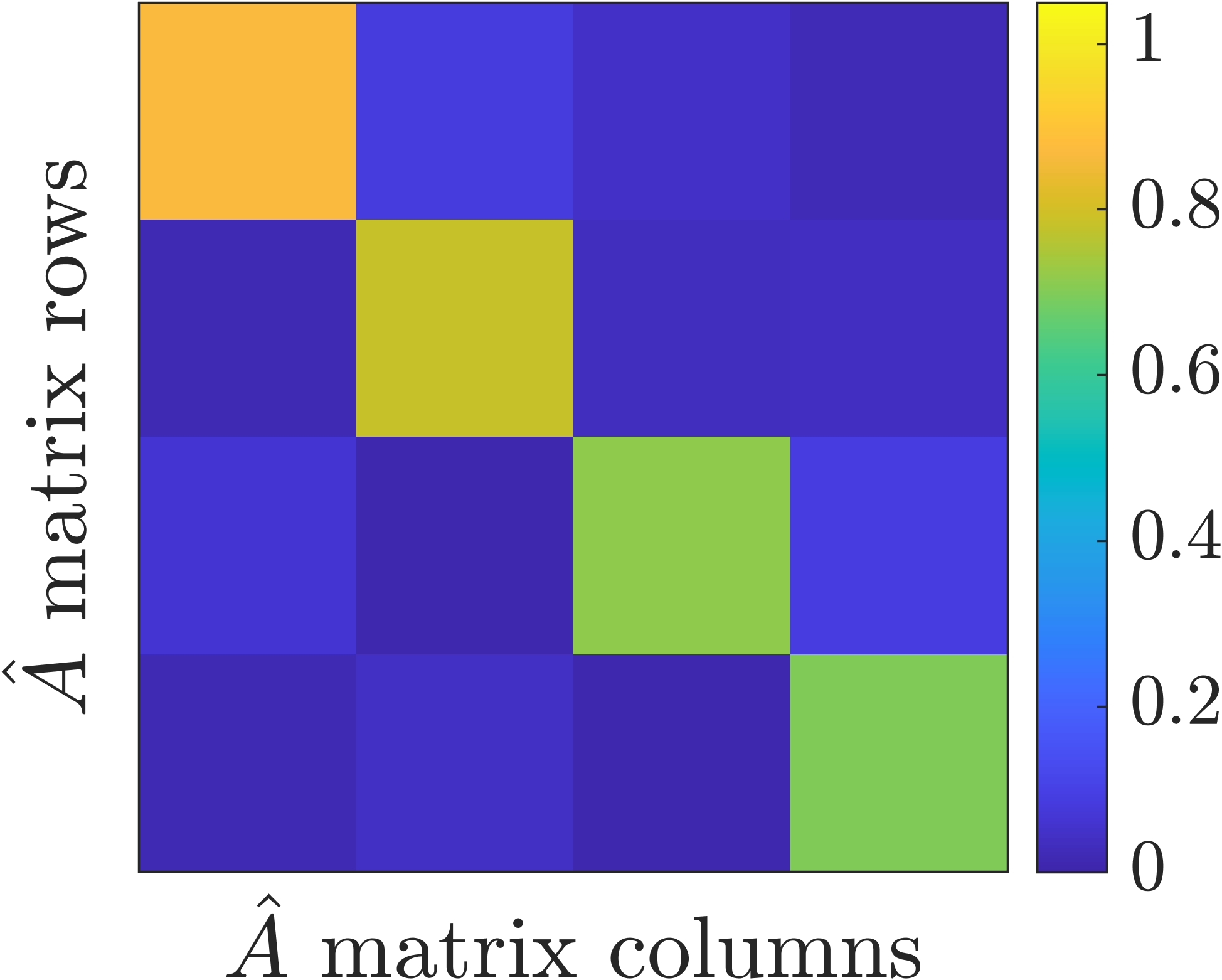}
         \caption{}
     \end{subfigure}
     \hspace{2mm}
     \begin{subfigure}{0.3\columnwidth}
    \includegraphics[width=\columnwidth]{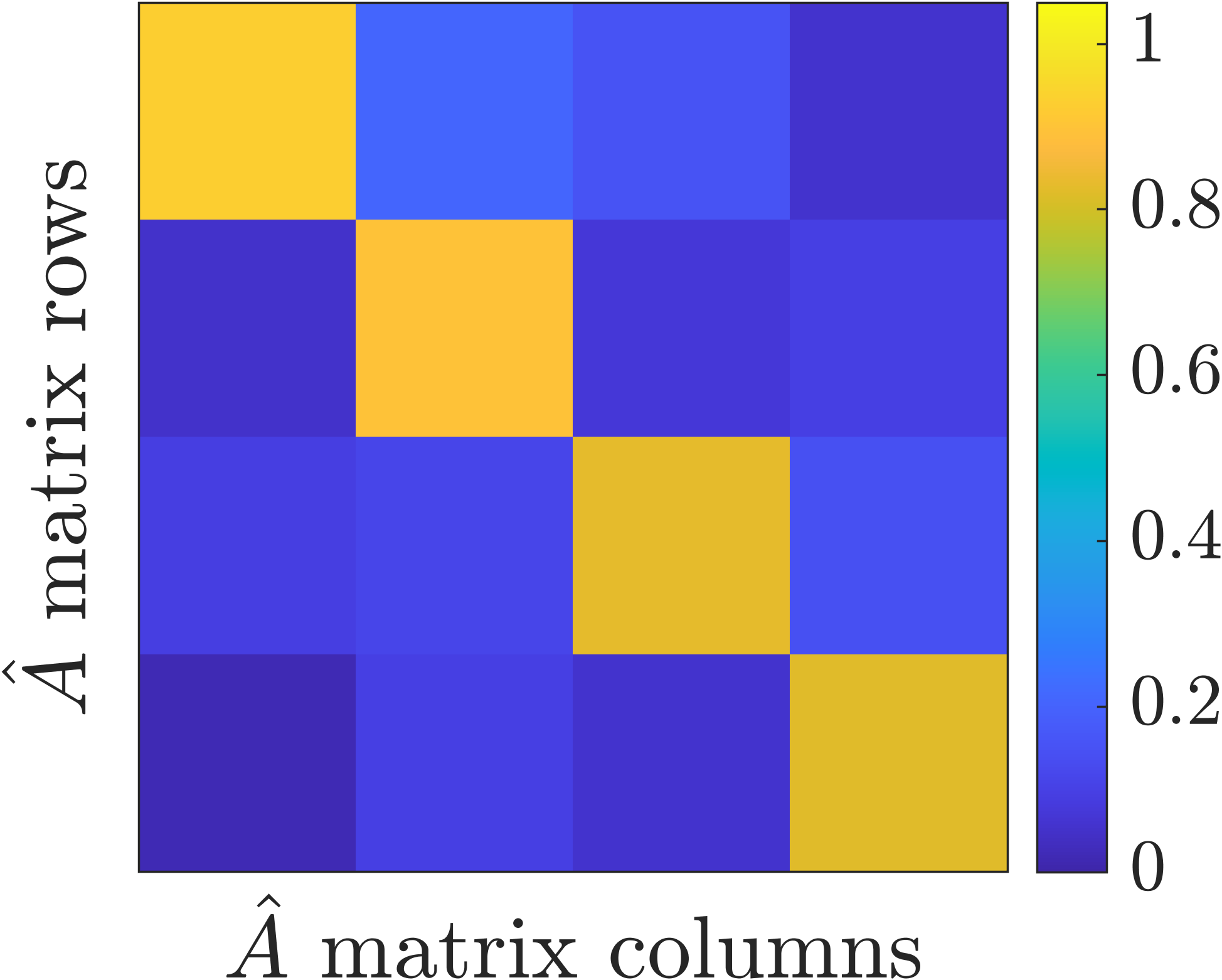}
         \caption{}
     \end{subfigure}
    \caption{Absolute values $ |\hat{A}|$ as per colorbar: a) least-squares solution after 1 iteration; b) after 100 iterations of Algorithm~1. Off-diagonal terms depict inter-channel connectivity.}
    \label{figureResA}
\end{figure}

We use $\rho = 1$ and $\lambda = 0$ and iterate 100 times through Algorithm~1. Results are shown in Figs.~\ref{figureResSeiz}  and \ref{figureResA}:
The algorithm not only denoises the time series, but also finds a higher connectivity estimate than the standard, least squares estimate without denoising (again, this is effectively the resulting $\hat{A}$ matrix after one iteration).

\subsection{NAR EEG modelling}

\begin{figure}[]
      \centering
      \includegraphics[width=0.6\columnwidth]{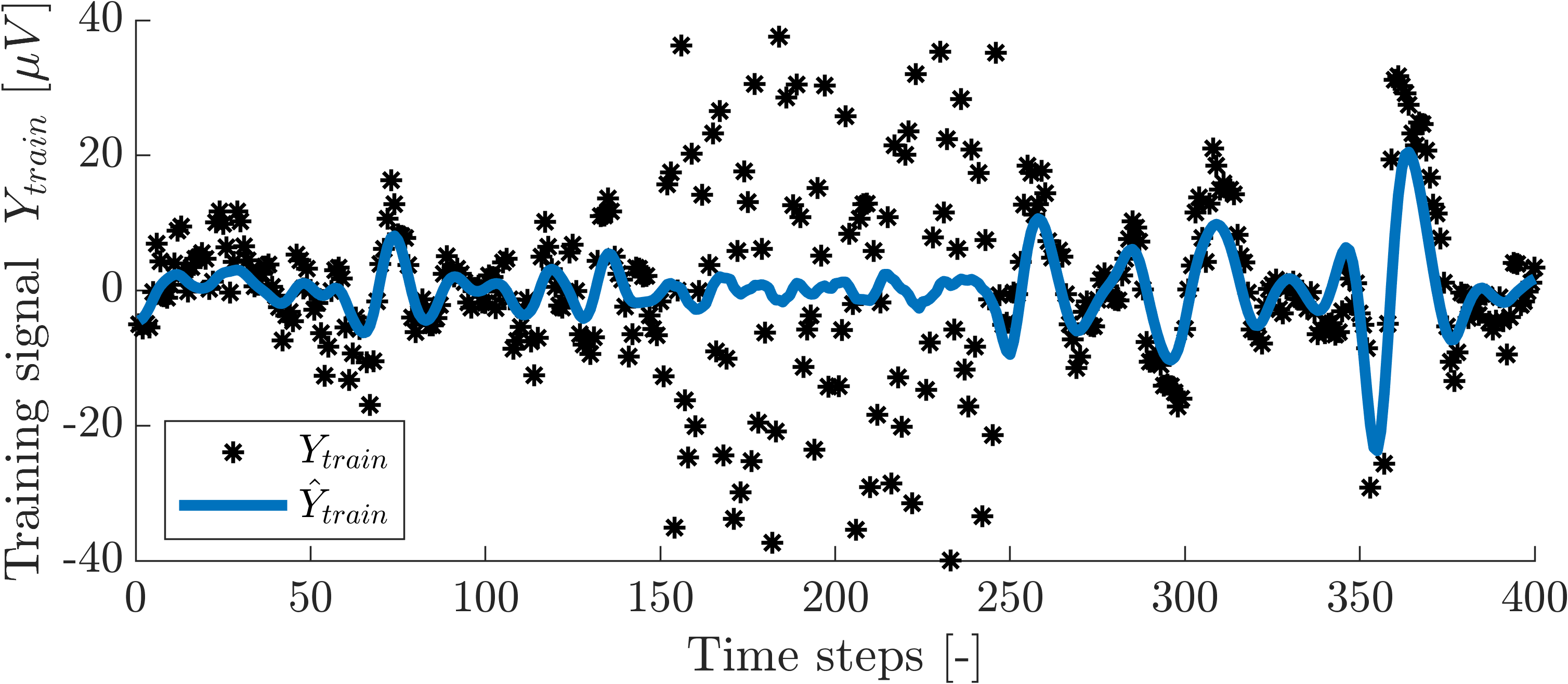}
      \caption{First-order difference in EEG channel~1 with random artefact, $Y_{\mathrm{train}}$ (black stars), and the denoised time series estimate, $\hat{Y}_{\mathrm{train}}$ (blue line), after 20 iterations of Algorithm~2.}
      \label{figureResSig1}
\end{figure}

\begin{figure}[]
      \centering
      \includegraphics[width=0.6\columnwidth]{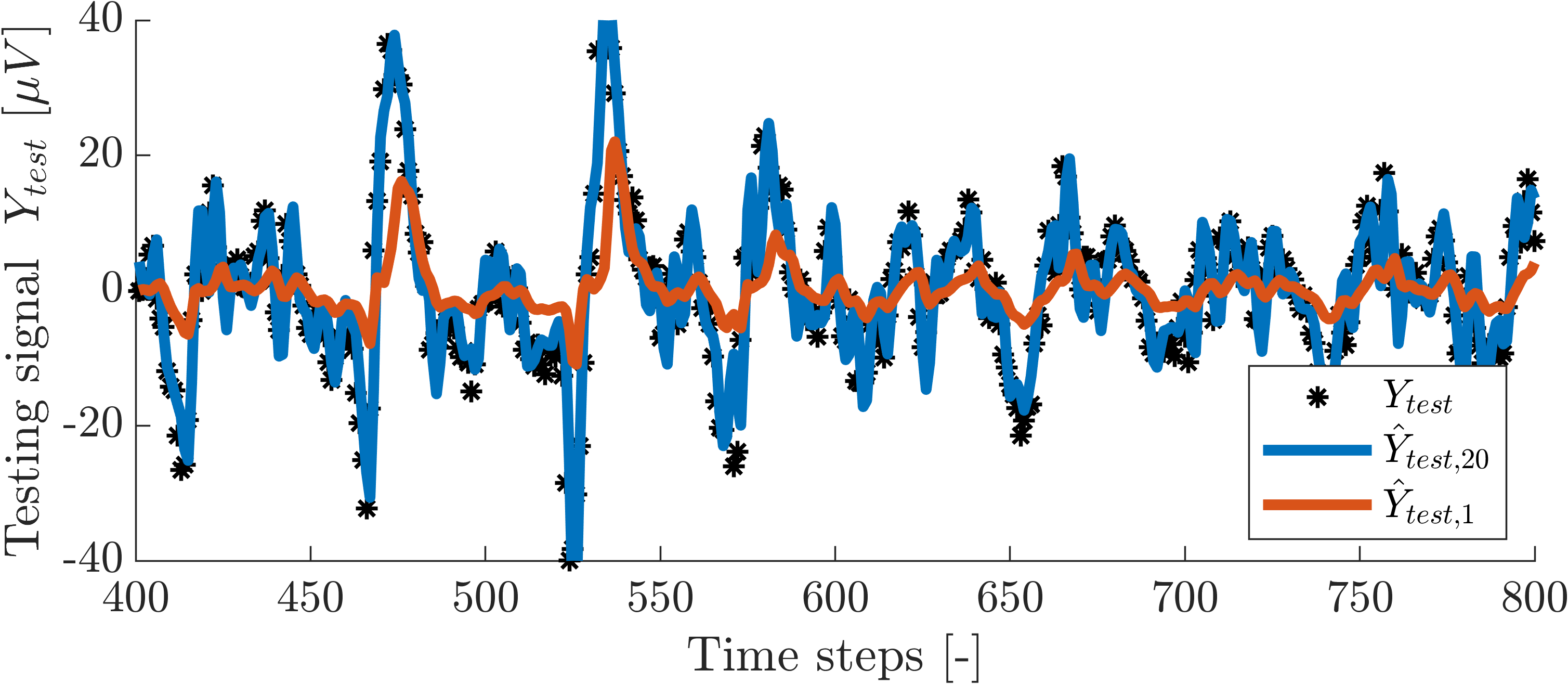}
      \caption{Plot of measurements $Y_{\mathrm{test}}$ (black stars) against the one-step-ahead predictions $\hat{Y}_{\mathrm{test},1}$ and $\hat{Y}_{\mathrm{test},20}$, after 1 (red) and 20 (blue) iterations of Algorithm~2, respectively.}
      \label{figureResSig2}
\end{figure}

We demonstrate a NAR case by modelling the EEG from Example \ref{example3} with the signature approach in Algorithm~2 and an artificially introduced artefact in the data.
To this end, we \textit{replace} the EEG in channel~1 of seizure~2 from time steps 150 to 250 ($0.25$ seconds) with random white noise with standard deviation of 20~$\mu V$. 
We then use the same first-order finite difference of the signal, $Y_{\mathrm{train}}$, for modelling, corresponding to the first 400 time steps with artefact.
We use $Y_{\mathrm{test}}$, corresponding to the second 400 time steps without artificial artefacts, for a test of the model. 
See Figs.~\ref{figureResSig1} and \ref{figureResSig2}. The task is to find a good model from $Y_{\mathrm{train}}$, despite the artefact, and measure its performance on predictions in $Y_{\mathrm{test}}$.

We use the signature implementation \textit{esig} \cite{Lyons2014} in Python, and parameters $r=4, k=2$, and depth $d = 2$. 
Note that higher embedding dimensions $r$ at fixed $d$ lead here to a more conservative denoising.
To calculate $s_t^d \in R^7$, we use the geometric path in $R^{4 \times 2}$ with $\hat{x}_t$ in the first column, and its cumulative sum $(\hat{y}_{t-r+1}, \hat{y}_{t-r+1} + \hat{y}_{t-r+2}, \cdots)^T$ in the second.
We apply Algorithm~2 for 20 steps with $\rho = 0.1$ and a small regularisation $\lambda = 0.001$ on states and parameters. 

First, we show that Algorithm~2 denoises the training data, i.e., ignores the artefact successfully (Fig.~\ref{figureResSig1}). 
Second, to elucidate the ability of the algorithm to find a reasonable model $\hat{A},\hat{C}$, we run the model on the test set $Y_{\mathrm{test}}$ in a one-step-ahead prediction: $\hat{y}_{t+1} = \hat{C} \hat{A} \hat{s_t}, ~ t=400, \cdots, 799$.
We illustrate the differences in $\hat{A}$ by plotting these predictions against the true $Y_{\mathrm{test}}$, for the first and last iteration of Algorithm~2 (Fig.~\ref{figureResSig2}). 
The first iteration, i.e., the classical least squares estimate for $\hat{A}, \hat{C}$ retrieved from signatures of the original signal $Y_{\mathrm{train}}$  with artefact, has a mean-squared-error of $e_{1} =  82~\mu V$. 
The model in the last iteration, i.e., the model from signatures calculated on the denoised signal shown in Fig.~\ref{figureResSig1}, has a mean-squared-error of $e_{20} =  31~\mu V$.

\newpage
\section{Discussion and Conclusion}

We have identified a procedure for estimating autoregressive models under measurement and state transition noise, as is typically encountered in biomedical signal processing. 
Assumptions on both these sources of uncertainty lead to a loss function that can be optimised via an iterative algorithm alternating between estimating model parameters and model states.
%The idea of using iterated steps to estimate model parameters and model states is by no means new. 
The described method is essentially an incarnation of the expectation-maximisation algorithm.
Our assumptions on the model type, i.e., autoregressive models, made it possible to derive analytic expressions for both the parameter and state estimation steps in a novel way. 
This alleviates substantial complexity, since this batch estimation requires no gradient-based optimisation, filtering or sampling techniques.  

Examples depicting the theoretical properties as well as performance on real-world EEG data were presented and analysed. 
Our results indicate that the proposed methodology is able to denoise time series and recover autoregressive coefficients accurately.
%Convergence properties can be deduced by realising the analogy to popular Expectation-Maximisation Algorithms. 
This might for example provide new ways to analyse the connectivity of brain regions under the inherent sources of noise in EEG data.

Another advantage of our procedure is that it works for linear and nonlinear autoregressive models alike, as long as the parameter estimation step itself is a linear problem. 
%The assumption of a linear parameter estimation step might seem like a limitation. 
%Otherwise, other optimisation algorithms may be applied for the presented loss function, such as gradient descent algorithms for neural networks.
We note that a very broad class of autoregressive models exhibit this property, and have shown an approach using path signatures as basis functions to model arbitrary nonlinear systems.
This makes the presented methodology widely applicable to tasks in biomedical signal processing and beyond.

%%%%%%%%%%%%%%%%%%%%%%%%%%%%%%%%%%%%%%%%%%%%%%%%%%%%%%%%%%%%%%%%%%%%%%%%%%%%%%%%
%\section{Acknowledgment}

\bibliographystyle{ieeetr}  
\bibliography{references}

\end{document}